# Simulating pharmaceutical treatment effects on osteoporosis via a bone remodeling algorithm targeting hypermineralized sites

MILAN Jean-Louis, CHAN YONE Claudia, ROSSI Jean-Marie, CHABRAND Patrick

Aix Marseille Univ, CNRS, ISM, Marseille, France

Abstract: Pharmaceutical treatments can slow bone degradation, thus reducing the fracture risk inherent in osteoporosis. Antiresorptive treatments block the over-activation of osteoclasts vs osteoblasts, but the resulting decrease in bone remodeling frequency may weaken bone structure over time, with no gain in bone volume. Anabolic treatments, however, induce gain in bone volume. The quantitative results from existing studies on the effects of treatments over time are general and non-patient-specific, while numerical models simulating evolution of patient-specific bone microarchitecture consider a spatially random distribution of the remodeling process. Here, we propose a new approach to simulate the remodeling over decades of an individual patient's bone microarchitecture, based on the hypothesis that the oldest sites, which are hypermineralized and more brittle, are remodeled first. Taking these older sites as prime targets of remodeling, simulations show that severe osteoporosis profoundly degrades the mechanical properties of the bone structure, which can be restored and even improved by anabolic, more than by antiresorptive, therapies.

# Abstract

Pharmaceutical treatments can slow bone degradation, thus reducing the fracture risk inherent in osteoporosis. Antiresorptive treatments block the over-activation of osteoclasts vs osteoblasts, but the resulting decrease in bone remodeling frequency may weaken bone structure over time, with no gain in bone volume. Anabolic treatments, however, induce gain in bone volume. The quantitative results from existing studies on the effects of treatments over time are general and non-patient-specific, while numerical models simulating evolution of patient-specific bone microarchitecture consider a spatially random distribution of the remodeling process. Here, we propose a new approach to simulate the remodeling over decades of an individual patient's bone microarchitecture, based on the hypothesis that the oldest sites, which are hypermineralized and more brittle, are remodeled first. Taking these older sites as prime targets of remodeling, simulations show that severe osteoporosis profoundly degrades the mechanical properties of the bone structure, which can be restored and even improved by anabolic, more than by antiresorptive, therapies.

## 1. Introduction

Osteoporosis is a major public health problem that reduces both the mass and the quality of bone, thereby increasing the risk of fracture, primarily in the elderly. Pharmaceutical treatments can slow bone degradation and thus reduce fracture risk. For instance, several clinical and experimental studies demonstrated the effects of antiresorptive treatments [1, 2, 3]. However, while these treatments block the over-activation of osteoclasts vs osteoblasts, less frequent bone remodeling may weaken bone structure in the long term, since antiresorptive treatments also lead to decreased bone repair. In contrast, anabolic treatments induce gain in bone volume and reduce fracture risk significantly [4, 5]. Yet the quantitative results on these treatments are based on averaged values and may not consider variability between subjects. Similarly, fracture risk is often assessed clinically using general scores based on bone densitometry, which may not fully predict the true mechanical strength of the patient's bone since patient-specific bone microarchitecture is not considered.

Imaging tools can provide precise 3D digital reconstruction of bone microarchitecture. Combined with mathematical algorithms of bone remodeling, these developments offer scientists and physicians the prospect of predicting the evolution of each patient's bone microarchitecture over decades, considering its individual pathological, genetic and epigenetic features. The numerical models developed to simulate bone remodeling can predict the evolution of real trabecular microarchitectures, with results comparable to clinical observations [6, 7, 8, 9]. Nevertheless, these models use a random or homogeneous remodeling without targeting old and damaged tissues. Models of mechanical adaptation of bone have also been developed but cannot simulate changes in bone microarchitecture caused by aging or pathologies [10, 11, 12, 13, 14]. To simulate bone remodeling as part of the aging process, the features of aged bone microarchitecture and the alterations in the biological activities of bone cells need to be considered.

In addition to potentially resorbing the bone matrix, regardless of its quality, to maintain plasma calcium homeostasis [17], bone remodeling also targets the damaged bone to repair it. Many studies analyzed the relationship between the initiation of remodeling and the presence of microcracks [18, 19, 20, 21, 22]. High mechanical stress induces bone damage, locally destroying the lacunar-canalicular network and killing the osteocytes, which interrupts the signals they emit to the bordering cells protecting the bone surface from osteoclastic resorption [23, 24]. Osteocyte apoptosis activates the recruitment of osteoclasts, which target injured bone to remove it [22,25]. Similarly, repeated mechanical stress induces bone fatigue cracks which initiate bone remodeling [17,26, 27].

Several in vivo studies observed that bone microcracks colocalize with hypermineralized areas [28,29]. Hypermineralized areas are the results of the second mineralization process that extends beyond the end of the apposition phase: when the normal quantity of minerals is deposited, crystals may continue to form, incrusting themselves in the matrix and replacing the interstitial water. While bone strength is normally positively correlated with bone mineral content, the hypermineralized bone becomes more brittle with higher ultimate stress and Young's modulus but lower ultimate strain, fracture work and thus toughness [21, 32, 33]. As a result, the hypermineralized bone is more likely to accumulate microcracks and fracture under repeated loads. Moreover, it has been reported that large linear microcracks occur more in hypermineralized areas [34,35] and induce more remodeling than diffuse microcracks [36]. Hypermineralization is also typical of the aged bone matrix and it is generally accepted that old bone is primarily remodeled by targeted bone remodeling. Aged bone has a longer history of significant mechanical loading, with accumulated damage. Moreover, the aging of the bone collagen matrix involves the accumulation of non-enzymatic, glycation-induced crosslinks, which decreases the structural ductility and toughness of bone material [37]. Studies have shown that resorption by osteoclasts increases in aged bone [38] and that osteoclast differentiation increases with increasing mineral content of the bone matrix [39, 40, 41]. Hypermineralized bone areas can thus be considered as prime targets of remodeling. They need to be considered in fracture risk prediction and when simulating the remodeling process [15; 16].

In the present study, we propose a new approach to simulate bone remodeling based on the hypothesis that the hypermineralized sites are remodeled first. Importantly, the study considers the secondary mineralization of deposited bone after the remodeling cycle. The objectives of this study were to simulate the aging of human trabecular architecture in both normal and osteoporotic cases and to analyze the benefits of pharmaceutical treatments based on either antiresorptive or anabolic agents.

## 2. Methods

### a) Cancellous Bone sample

We present hereunder the source and properties of the sample of human trabecular bone used as input for the microarchitecture remodeling model. This study used one femoral head from an 87-year-old woman undergoing surgery for femoral neck fracture (T-score of -2.1, DXA exam) in the Orthopedic

Surgery Department of the Sainte Marguerite Hospital in Marseille, France. With a T-score of -2.1, the patient was diagnosed with osteopenia. Informed consent was obtained from the patient before treatment. A cubic specimen (10x10x10mm$^3$) was extracted from the femoral head and the marrow was removed in a water bath (37°C) containing a stirring rod. Any remaining marrow was extracted with compressed air. The sample was then scanned using high resolution µCT tomography (Skyscan® 1172, Kontich, Belgium) to obtain a trabecular bone image with an isotropic voxel size of 20 µm. This system operates via the cone-beam method and comprises a sealed microfocus X-ray tube with a spot size less than 8 µm and an 11 Mp 12-bit CCD camera. Projection images were obtained at 80 kV and 100 µA with a 0.5 mm aluminum filter. Cross-section images were stored in 16-bit format (1000x1000 pixels). A phantom made of hydroxyapatite with a density of 1.75g/cm$^3$ was placed under the sample. The phantom represented a piece of cortical bone with a density of 2.17g/cm$^3$, defined as the mass divided by the bulk volume, as well as the mineral content of normal mineralized bone tissue either from cortical or from cancellous bone, whose mineral content is similar at microscale [42]. The phantom gave a gray-value reference, which was used to identify local material components and the related mechanical properties. µCT scanning revealed that the bone sample had a Bone Volume / Total Volume (BV/TV) of 22.18%, a Bone Surface/Bone Volume (BS/BV) of 10.18 mm$^{-1}$, a Trabecular Thickness (Tb.Th) of 0.204mm, a Trabecular Number (Tb.N) of 1.087 mm$^{-1}$, a Trabecular Separation (Tb.S) of 0.769 mm, and a Degree of Anisotropy of 0.677 calculated as 1 - smallest eigenvalue / largest eigenvalue [43; 44]. In the 3D segmented trabecular bone microarchitecture, 70% of voxels had a gray-value above that of the cortical phantom (Supplementary Data Figure 1). To simulate the remodeling process, a central cubic sub-volume equivalent to 4x4x4 mm$^3$ (200x200x200 voxels) was analyzed. A coarsening method [45, 46] was applied to reduce both the number of voxels and the number of degrees of freedom for finite element analysis, by grouping the microCT voxels of 20x20x20µm$^3$ into larger voxels of various sizes, which were selected using a sensitivity study. The gray-value of the new voxels, herein termed coarsened voxels, was the average of the gray-value of the initial voxels. This method could, however, remove bone material heterogeneities, masking the extent of the hypermineralization due to the remodeling process. We therefore conducted a mesh sensitivity study on the coarsening method with several voxel sizes: 20, 40, 60, 80, 100µm. Results indicated, for instance, that most hypermineralized areas disappeared in a mesh with voxels as large as 100µm. We thus identified coarsened voxels of 60µm as the best compromise between computational cost and the preservation of bone material heterogeneities (Supplementary Data Figure 2). For the rest of the manuscript, the term *voxels* refers to these 60µm coarsened voxels. In this configuration, 66.7% of voxels had a gray-value above that of the cortical phantom.

Finite element models were generated directly by converting voxels to equally-sized linear eight-node brick elements. For all models, the element material properties were assumed to be isotropic linear elastic. Poisson's ratio was set to 0.3 and the Young's modulus value of each element e, $E^e$, was calculated using the power-law Equation (1) proposed by previous authors [45]:

$$E^e = E_{max} * (GV^e)^\gamma \qquad (1)$$

where $GV^e$ is the normalized gray-value of the element and γ an exponent set to 1.5 [45]. $E_{max}$ is determined via Equation 1 applied to normalized cortical bone, such that $GV^e = GV_{cort}$, with $GV_{cort}$ the normalized cortical phantom gray-value and $E^e = E_{cort}$ with $E_{cort}$ the elastic modulus of cortical bone defined by nano-indentation and set to 19 GPa [42, 47].

The Young's modulus distribution over the whole set of voxel elements followed a Gaussian distribution centered at 23 GPa and ranged from 5 GPa to 40 GPa for 97% of elements. This range of tissue stiffness is consistent with the literature [48].

### b) Bone remodeling algorithm

The bone remodeling algorithm we propose is an iterative process repeating full 6-month remodeling cycles. Bone resorption and deposition are numerically modeled respectively by deleting or adding voxels in the bone microarchitecture. Mineralization is modeled by the stiffening of the bone matrix over time. The main variables of the model are the total remodeling surface area of the bone microarchitecture, RA, and the gain or loss in bone volume after remodeling, δ. RA reflects the frequency of basic multicellular unit (BMU) activation: the greater the RA, the more osteoclasts are recruited. For instance, in normal aging, RA is equal to 7% because at any time, 7% of the trabecular bone surface is undergoing a 6–month cycle of remodeling [10]. RA is divided into N resorption pit sites of 300µm-diameter. δ represents the difference in activity between osteoblasts and osteoclasts. δ quantifies at each resorption site the loss (-100%<δ<0) or gain (0<δ<100%) in bone volume : in a resorption pit, δ=0 means that osteoblasts deposit the same volume of osteoid tissue as that resorbed; δ=-100% means that no tissue is deposited after resorption; and δ=100% means that the volume of osteoid tissue deposited is twice that resorbed. For instance, in normal aging, δ is equal to -5% [10]. We consider that the new bone deposited is mature and mineralized and that its Young's Modulus is equal to $E_{cort}$. We also include the parameter $K_{hyper}$ in GPa that governs the increase in the Young's modulus of the local bone matrix due to a secondary mineralization potentially leading to hypermineralization. $K_{hyper}$ is identified numerically via a parametric study by assessing the proportion of hypermineralized elements, denoted α, before and after 10 years of remodeling over the whole bone sample. Hypermineralized elements possess an elastic modulus E greater than $E_{cort}$. By modifying RA and δ, different remodeling scenarios can be simulated. Increasing RA and/or |δ| (δ <0) can simulate a case of menopause, where the frequency of remodeling activation and the imbalance between osteoclasts and osteoblasts have been shown to increase simultaneously [49].

Because our aim was to explore only how the biological imbalance between osteoclastic and osteoblastic activities might affect bone remodeling, we considered the mechanical loading as constant and did not incorporate it into the bone algorithm proposed here.

The successive steps of the bone remodeling algorithm are presented hereunder

**Activation of bone remodeling**

The N resorption pits were identified on the surface according to the degree of mineralization of the tissue underneath, characterized by its local Young's modulus. The most mineralized sites were remodeled first.

**Resorption, deposition and mineralization of bone matrix**

To represent the typical dimension of 300µm-diameter hemispherical resorption pit in the trabecular bone model composed of 60µm voxel elements, we considered a resorption pit 5 elements wide and 2 elements deep, centered on the remodeling site. All the elements circumscribed by the resorption pit were modified by the remodeling process. The remodeling algorithm was integrated into a finite element code, where each iteration corresponds to a full 6-month remodeling cycle [30]. At each cycle, tissue is resorbed, and then new bone tissue is deposited with new elastic properties after primary mineralization. Osteoclastic and osteoblastic activities may not balance each other, and consequently bone mass can decrease or increase. In the model, at a remodeling site, elements may be suppressed or created depending on the value of bone imbalance δ (Supplementary Data Figure 3). If δ = -100%, two layers of elements are suppressed. If δ = -50%, only one layer of elements is suppressed, while the remaining elements of the second layer are given Young's modulus of cortical bone $E_{cort}$. If δ = 0%, then, no element is created or suppressed, and element stiffness is set to $E_{cort}$. If δ = 50%, one layer of new elements with $E_{cort}$ is created at the surface of the remodeling site, while δ = 100% leads to the creation of two layers of new elements with $E_{cort}$. When δ takes intermediate values such as -5%, -10%, -25% or 25% as in some cases of our study which are described in the following sections, the volume of tissue to be removed or deposited does not fit a whole number of element layers. In such cases, bone imbalance is performed by adapting ρ, the density of bone material, in elements of the top layer of the remodeling site. These elements with ρ < 1 are given an effective Young's modulus E according to a mixing law between void and cortical bone ($E_{cort}$). For instance, if δ =-25%, elements of the lower layer at a remodeling site are given E = $E_{cort}$ while elements of the top layer are given ρ = 0.5 and E = 9.5 GPa. If this site was remodeled again immediately, without additional mineralization, with the same bone imbalance of δ =-25%, elements of top layer would be given ρ = 0.125 and E = 2.4 GPa (Supplementary Data Figure 3g and h). In general, when bone imbalance is negative, this formulation decreases local rigidities. When the element's Young's modulus is below the threshold value $E_{min}$=0.1 GPa, the element is deleted.

**Secondary mineralization of bone matrix.**

Since the aging of bone tissue involves locally increased quantities of minerals that stiffen bone material, aging was modeled here by increasing the local Young's modulus, while the Poisson's coefficient remained constant at 0.3. In the model, at each cycle, all elements except those just deposited stiffen as their mineral content increases over time. To estimate the stiffening rate during secondary mineralization, we combined an expression of local Young's modulus of femur trabecular bone with ash density [50] and an expression of ash density with time during secondary mineralization [24]. By combining the two expressions, we obtained a very slow evolution of Young's modulus with time during secondary mineralization, which can be simplified as an affine function with a slope equal to $K_{hyper}$ ~ 0.5 GPa/year. The Young's modulus of element e at iteration i +1, $E^e_{i+1}$, is given by Equation (2)

$$E^e_{i+1} = E^e_i + K_{hyper} \qquad (2)$$

The parameter $K_{hyper}$ controls the rate of bone material stiffening over time, and thus the rate of bone material aging over time. We imposed a maximum stiffness limit: if $E^e_{i+1} \geq E_{max}$, then $E^e_{i+1} = E_{max}$, with $E_{max}$ = 50GPa, corresponding to the maximum value identified in the μCT images. Parametric analysis was performed to identify a consistent value of $K_{hyper}$. Figure 1 shows, for different $K_{hyper}$ values, the distribution of the elements as a function of their Young's modulus values after 10 years of bone remodeling. When $K_{hyper}$ = 0.1GPa, there is a maximum of elements whose Young's modulus is between 15 and 19 GPa, and therefore $K_{hyper}$ is not high enough to induce hypermineralization over time. By contrast, when $K_{hyper}$ = 1GPa, there is a maximum of elements whose Young's modulus is greater than 40 GPa and the mineralization value appears too high. When $K_{hyper}$ = 0.5GPa, there are far more elements with a Young's modulus greater than 35 GPa than before remodeling. Finally, for $K_{hyper}$ = 0.3GPa, most elements have a Young's modulus between 15 and 25 GPa and a difference from the initial architecture can therefore be observed. The proportion of hypermineralized elements (E ≥ $E_{cort}$), denoted α, is 66.7% before remodeling. When the different configurations are compared, after 10 years of remodeling, with $K_{hyper}$ = 0.3GPa the proportion of hypermineralized elements is slightly greater (α = 69.0%) than before remodeling. This seems relevant to clinical practice, as a slight increase in mineralization is clinically observed with age. $K_{hyper}$ = 0.3GPa was therefore chosen for use in the remainder of the study.

### c) Bone remodeling patterns

**Normal aging case**

To model a normal remodeling process, RA and δ were given values from the literature, respectively 7% and -5% [10] (Table 1).

**Osteoporotic case**

The osteoporotic case was simulated by doubling the RA and/or δ of the normal aging case [49]. We considered 3 osteoporotic patterns: (RA=14%, δ=-5%); (RA=7%, δ=-10%); (RA=14%, δ=-10%), the last expected to be the most severe (Table 1).

**Antiresorptive treatment**

Treatment for osteoporosis may involve antiresorptives, also called anticatabolics, which make it possible to block remodeling and thus limit bone microarchitecture degradation. This treatment was simulated by using (RA=3.5%, δ=-5%) (Table 1).

**Anabolic treatment**

Osteoporosis can also be treated by anabolic agents such as parathormone, which stimulate bone formation. To simulate anabolic treatment, it was assumed here that excess bone tissue is deposited at each cycle of remodeling, leading to a positive bone balance δ > 0, i.e. over-activity of the osteoblasts compared to the osteoclasts (Table 1). This treatment was simulated by using (RA=7%, δ=25%), (RA=7%, δ=50%) or (RA=7%, δ=100%). Apposition results in the creation of new elements, as detailed above.

**Effect on bone microarchitecture stiffness**

To estimate over time the impact of bone remodeling on the mechanical behavior of bone microarchitecture, we simulated successively, for each bone remodeling year, compression tests on the trabecular bone sample subjected to all cases of bone remodeling: normal or pathological aging, with or without treatment. Based on finite element method, the compression test consisted of imposing -0.15% of deformation on the bone sample. The compression tests were simulated successively in the 3 x-y-z directions. By comparing the global stress and strain levels, we calculated the values of the apparent Young's modulus of the trabecular bone sample in the 3 directions and in all the cases of bone remodeling.

## 3. Results

### a) Normal aging case

Simulation of bone remodeling over 20 years in the normal aging case shows a thinning of trabeculae (Figure 2b compared with Figure 2a) with the turnover of the most mineralized sites (Figure 3). Aging of the tissue resulting from secondary mineralization is also observed, as well as a degradation of the microarchitecture, with some gaps on the surface. The simulation indicated a slight decrease in BV/TV (Figure 4). Note that total volume (TV) did not change over time, and that relative change in BV/TV means relative change in BV. Finite element analyses indicated that the apparent Young's modulus of the sample decreased during the first decade by 18% on average, over the 3 x-y-z directions (Figure 5).

### b) Osteoporotic case

A more important decrease in BV/TV was obtained by doubling one or both of (RA, δ) compared to the normal case (Figure 4). An identical effect on bone volume was obtained by doubling either RA or δ. However, doubling both RA and δ did not double the effect obtained by doubling only one of the two parameters. The bone geometry obtained after 20 years of remodeling by using (RA=14%; δ=-5%) and (RA=14%; δ=-10%) is displayed respectively in Figures 2c and d. This can be compared to the initial bone structure and its normal case evolution (RA=7%; δ=-5%) displayed respectively in Figures 2a and b. The mode of degradation of the bone microarchitecture over time under the most severe degree of osteoporosis (RA=14%; δ=-10%) is displayed in Figure 6, showing the typical bone degradation mechanisms: plate perforation, thinning of the trabeculae and disappearance of certain bone parts. Osteoporotic remodeling patterns decreased apparent bone stiffness over time. Indeed, the percentage decrease in apparent bone stiffness ranged between 27-49% after the first 5 years and between 40-66% after 10 years, depending on the osteoporotic case (Figure 5). As expected, this effect was more marked in the most severe pattern of osteoporosis when the parameters were doubled (RA = 14% and δ = -10%), particularly during the first 5 years.

### c) Antiresorptive treatments

Simulation results indicate that, over a 5-year period, antiresorptive treatment slowed the decrease in BV/TV after 5 years of treatments (Figure 7). Although antiresorptives did not here allow a net gain in bone mass and only reduced the amount of bone loss, they completely corrected the effect of osteoporosis, leading to even less degradation in bone quantity than with normal aging. Compared with the severe osteoporotic case (RA = 14% and δ = -10%), antiresorptive treatment led to a higher BV/TV with relative difference of 2.4 % after 5 years (Figures 4 and 7). However, under antiresorptive treatment, the remodeling rate slowed and the proportion of hypermineralized elements (E ≥ $E_{cort}$) was greater than in the untreated case (Figure 8). Antiresorptive treatment slightly decreased the apparent Young's modulus, by 5%, after 5 years (results not shown). The model did not indicate a net gain in stiffness for the overall bone structure compared to the initial state. This means that the increase in local Young's modulus due to secondary mineralization did not balance out the bone loss. Nonetheless, compared with untreated severe osteoporosis, antiresorptives led to higher apparent stiffness of bone with relative difference of 86% after 5 years.

### d) Anabolic treatment

With a remodeling surface (RA) of 7% and positive δ values, the results of the simulations indicate an increase in BV/TV after 5 years of treatment (Figure 9). Greater increases in bone volume appeared when the RA value was raised, which amounts to increasing remodeling frequency. The smoothed and voxel microarchitectures (finite element model) after 5 years of treatment are shown in Figures 10 and 11. Many trabecular plates thickened. The macro equivalent stiffness of the whole bone structure increased with time (Figure 12). A great difference in mechanical stiffness of bone was observed between case (RA=7%; δ=-5%) and case (RA=7%; δ=25%). However, there was only a slight difference in mechanical behavior between δ = 25% and δ = 50%. It appears that, above a certain value, increasing δ does not greatly impact the apparent stiffness, but does lead to an increase in BV/TV ratio (Figure 9).

## 4. Discussion

### a) Simulated effect of aging on bone remodeling

To simulate bone remodeling in the normal aging case, we considered that 7% of the surface is remodeled every 0.5 years, the lacuna of resorption is a 0.150mm-radius half sphere, and osteoblasts deposit 95% of the volume resorbed. Applying this remodeling pattern to a cancellous bone sample with a ratio of BS/BV=10.18mm$^{-1}$ led to a rate of Bone Formation/Bone Volume/year (BF/BV/y) equal to 13.5% as calculated in Supplementary Data Figure 4. This value is consistent with the mean value of this index reported by Recker et al. in 1988 [51]. For the normal aging values of parameters RA and δ, the simulation indicated a decrease in BV/TV of 1% per decade. This result is consistent with other independent studies from the literature. Recker et al. found a decrease of BV/TV per decade of 1.8 % in 30 post-menopause women [51]. Riggs et al showed equivalent bone loss with a decrease in BV/TV of 2.3% per decade in 700 male or female subjects aged from 20-98 years [52]. Yuen et al. in 2010 showed in 1300 male or female subjects aged from 20-98 years a variation in BV/TV per decade of 3.7% [53]. MacDonald et al. studied trabecular bone microarchitecture in 644 Canadian adults and estimated an equivalent decrease of 20-26% in BV/TV over 7 decades in subjects from 20 to 90 years old [54]. Based on this literature, the bone remodeling pattern of normal aging should be close to what we considered as the osteoporotic patterns (RA=7%; δ=-10%) or (RA=14%; δ=-5%), leading to a similar decrease of 2.4% per decade in BV/TV.

This first simple approach, based on the assumption of preferential old bone turnover, makes it possible to simulate the long-term remodeling of a human bone. By contrast with the models previously developed [6, 7, 8], our simulation of remodeling is not random. The typical bone degradation mechanisms, like trabecular plate perforation, thinning of spans and finally total disappearance of certain bone parts, are demonstrated (Figs. 9 and 10). Mechanical properties decrease with time, indicating the weakening of the trabecular microarchitecture (Figure 5).

In the osteoporotic case simulations, BV/TV decreased by 2-5% per decade. These results are consistent with other studies. For instance, Chen et al. 2013 combined several results from different studies on aged and osteoporotic trabecular bone microarchitecture and found a decrease of 22% in BV/TV over 3 decades from the age of 60 to 90 in trabecular bone in vertebra and femur neck [55]. Mazess estimated an equivalent decrease of 6-8% per decade in BV/TV [56].

Our study indicates that osteoporotic patterns decreased the average apparent Young's modulus of the bone sample by 30-49% during the first 5 years and by 40-66% after 10 years. These levels of alteration in stiffness are consistent with findings from studies using ovariectomized (OVX) rats as an in vivo model of osteoporosis [57,58]. In these studies, Kostenuik et al in 2001 and Shahnazari et al. in 2011 tested rat lumbar vertebra trabecular bone under compressive loading and showed respectively decreases of 47% and 32 % in apparent Young's modulus when the OVX rats were compared with a sham group.

### b) Simulated effects of antiresorptive treatment

Our results indicate that antiresorptives slowed BV/TV decrease to -0.3% after 5 years and increased apparent stiffness: by 30% compared to osteoporotic patterns (RA=14%; δ=-5%) and (RA=7%; δ=-10%) and by 86% compared with the most severe osteoporosis pattern (RA=14%; δ=-10%). BV/TV maintenance was observed in clinical studies on postmenopausal women treated for one year with alendronate (ALN) [59] or with risedronate [60]. Our results are consistent with Khajuria et al. 2014, who reported an increase of 75-87% in apparent Young's modulus of lumbar vertebra tested under compression in OVX rats treated with antiresorptive zoledronic acid compared with untreated OVX rats [61]. Shahnazari et al. in 2011 found that 4-month antiresoptive treatments with alendronate (ALN) and with raloxifene (RAL) increased by 12% and 30%, respectively, the apparent Young's modulus in OVX rats [58]. In our simulations, an increase in Young's moduli after antiresorptive treatment suggested an increase in BMD. Using an equation from Keller et al. 1994, $E=10.5\ \rho^{2.29}$, the mineralization rate corresponding to K=0.3 GPa/0.5 years is 0.29g/cm$^3$/year [62]. Element density with E=19 GPa is 1.29 g/cm$^3$. Thus, the secondary mineralization simulated in the present study corresponds to an increase in BMD of 0.29/1.29=22%/year. Other studies from the literature measured an increase of 4.5 % in BMD in response to alendronate treatment [63]. They found a 4.5% increase after 1 year and an 8% increase after 2.5 years. The increase in local Young's modulus due to secondary mineralization and the low rate of bone remodeling we noted in the simulated antiresorptive pattern are consistent with the local stiffening of bone matrix that Cheng et al. in 2009 observed in old osteoporotic rats after a prolonged antiresorptive treatment [64]. Local increase in mineralization in pathological cases under antiresorptive treatments was observed by Boivin et al. (2000), who found that while there was no net gain in bone mass during an antiresorptive treatment with alendronate, mineralization increased [1]. Mashiba et al. (2001) and Li et al. 2001 showed an abnormal frequency of bone microcracks and a 20% average decrease in bone mechanical resistance in dog populations treated for 1 year with anticatabolic agents (risendronate, alendronate and bisphosphonate) at doses 6 times higher than in humans [3, 65]. Our study indicated that overly long treatments promote hypermineralization, which may impair the mechanical properties of the tissue.

### c) Simulated effects of anabolic treatment

Simulation of bone remodeling using the highest anabolic pattern (RA=7%; δ=100%) increased BV/TV by 2.5% after 5 years. Compared to the most severe osteoporosis case, this anabolic treatment increased BV/TV by 11%. These results agree with in vivo studies based on OVX rats which showed the same trend with even greater increase. For instance, Shahnazari et al. in 2011 showed that 4 months of anabolic parathyroid hormone (PTH) treatment increased the vertebral trabecular BV/TV of OVX animals by 53% [58]. Nonetheless these benefits are dependent on continuation of PTH treatment: after withdrawal of the treatment, PTH-treated OVX rats tended to have the same properties as OVX rats. As stopping PTH led to higher rates of bone remodeling, the authors suggested combining PTH with antiresorptives such as ALN, to preserve the benefits of PTH concerning bone volume. Since decreasing RA reproduces antiresorptive effects and increasing δ reproduces anabolic effects, our model could simulate sequential or combined therapies for anabolic and antiresorptive agents by using either sequential patterns of (RA; δ) or a mixed one.

Our simulations of anabolic treatment increased apparent Young's modulus by 11% compared to initial state and by 125% compared to the most severe osteoporosis. Our results are consistent with other studies reporting benefits from anabolic treatments. For instance, coupling FE analyses with μCT, Graeff et al. in 2009 estimated similar increases in bone vertebra stiffness: 20% after 1 year of anabolic teriparatide treatment and of 26% after 2 years of treatment [66]. Kostenuik et al showed that 5 months of PTH treatment increased apparent Young's modulus by 32% in OVX rats compared to untreated OVX rats [57].

Comparing our results with clinical studies might suggest that the anabolic pattern we considered was underestimated. Clinical study from Dempster et al in 2001 showed that 36-month anabolic PTH treatment led on osteoporotic women increased BV/TV about 5,2% compared to baseline [67]. Khastgir et al., (2001) in osteoporotic women treated with estrogen, showed a trabecular volume increase of 5.7% over 5 years [4]. Yet we were able to simulate this trend, twice as great, by simply increasing RA, which amounts to increasing remodeling frequency. Of the values tested, RA = 25% and δ = 100% were associated with a bone volume increase of 5.8% after 5 years of treatment. Moreover, the simulated trabecular thickening also matched that found experimentally by Khastgir et al., (2001) [4].

Our finite element analyses show here that the macroelastic properties of bone increase with time when it remodels under anabolic treatment. If the results of our model can be translated to real cases, the mechanical adaptation of trabecular bone tissue under anabolic treatments appears to rely on thickening of the spans and rigidification of the overall structure. This suggests that anabolic treatments increase the overall rigidity of the bone by modifying its microarchitecture (its volume and the arrangement of the spans). The results of this numerical study agree with clinical observations which show that anabolic agents stimulate bone formation and are effective in the treatment of osteoporosis [4]. However, this treatment's main limitation is its mode of administration, by subcutaneous auto-injection, and its cost. In France, anabolic treatment is covered by the state health insurance only for women who have already had 2 vertebral fractures [68], and therefore its use is reserved for severe forms of vertebral osteoporosis.

### d) Limitations and future perspectives

Our simple approach based on bone remodeling algorithm targeting hypermineralized sites can simulate a relevant temporal evolution of the bone microarchitecture of any patient and thus predict the structural consequences of a bone remodeling scheme corresponding to a given pharmaceutical treatment. In the case of an osteoporotic patient, the model can potentially contribute to clinical decision-making by predicting the long-term structural effect of pharmaceutical treatments. In a future study, it would be interesting to consider a wide range of bone microarchitectures, T-scores, BMD, BV/TV, trabecular thickness and separation, trabecular anisotropy from a panel of osteoporotic patients monitored clinically. This would validate the model by comparing the simulation results to the actual bone microarchitecture after treatment. Moreover, just as the present study aims to identify the influence of pharmaceutical treatments alone on the osteoporotic bone microarchitecture, future models could integrate mechanical loading as one of the main factors controlling bone remodeling to analyze the effect of physical activities in interaction with a given treatment..

# 5. Conclusion

We have developed a model of the evolution over decades of an individual patient's bone microarchitecture and simulate different treatment scenarios, evaluating their effects on mechanical stiffness of bone at both micro and macro scales. This simple and efficient contribution to existing numerical approaches for bone remodeling may be useful for clinical applications in the treatment of osteoporosis. In the future, routine radiological examination will likely detect the hypermineralized bone matrix, so this type of bone remodeling model targeting aged bone sites may be useful for evaluating the potential effect of medications on a patient's bone microarchitecture.

# 6. Acknowledgements

Competing interests: None declared

Funding: None

Ethical approval: Not required

# References

[1] Boivin GY, Chavassieux PM, Santora AC, Yates J & Meunier PJ. Alendronate increases bone strength by increasing the mean degree of mineralization of bone tissue in osteoporotic women. Bone 2000;27;5:687–694. https://doi.org/10.1016/s8756-3282(00)00376-8

[2] Turner CH. Biomechanics of bone : Determinants of skeletal fragility and bone quality. Osteoporosis International 2002;13;2:97–104. https://doi.org/10.1007/s001980200000

[3] Mashiba T, Turner CH, Hirano T, Forwood MR, Johnston CC & Burr DB. Effects of suppressed bone turnover by bisphosphonates on microdamage accumulation and biomechanical properties in clinically relevant skeletal sites in beagles. Bone 2001;28:524–531. https://doi.org/10.1016/s8756-3282(01)00414-8

[4] Khastgir G, Studd J, Holland N, Alaghband-Zadeh J, Fox, S & Chow J. Anabolic effect of estrogen replacement on bone in postmenopausal women with osteoporosis : Histomorphometric evidence in a longitudinal study. Journal of Clinical Endocrinology & Metabolism 2001;86;1:289–295. https://doi.org/10.1210/jcem.86.1.7161

[5] Meunier PJ. Traitement de l'ostéoporose postménopausique par les agents anaboliques. Revue du Rhumatisme 2001;68:944–950.

[6] Liu XS, Huang AH, Zhang XH, Sajda P, J, B. & Guo XE. Dynamic simulation of three dimensional architectural and mechanical alterations in human trabecular bone during menopause. Bone 2008;43:292–301. https://doi.org/10.1016/j.bone.2008.04.008

[7] Müller R. Long-term prediction of three-dimensional bone architecture in simulations of pre-, peri- and post-menopausal microstructural bone remodeling. Osteoporosis International 2005;16: S25–S35. https://doi.org/10.1007/s00198-004-1701-7

[8] Van Der Linden JC, Verhaar JAN, Pols HAP & Weinans H. A simulation model at trabecular level to predict effects of antiresorptive treatment after menopause. Calcified Tissue International 2003;73:537–544. https://doi.org/10.1007/s00223-002-2151-x

[9] Christen P, Ito K, Müller R, Rubin MR, Dempster DW, Bilezikian JP, Van Rietbergen B. Patient-specific bone modelling and remodelling simulation of hypoparathyroidism based on human iliac crest biopsies. J Biomech 2012;45;14:2411–2416. https://doi.org/10.1016/j.jbiomech.2012.06.031

[10] Cowin SC. Bone Mechanics Handbook . Boca Ranton 2001.

[11] Tsubota K, Suzuki Y, Yamada, T, Hojo M, Makinouchi, A. & Adachi, T. Computer simulation of trabecular remodeling in human proximal femur using large-scale voxel FE models : Approach to understanding Wolff's law. Journal of Biomechanics 2009;42;8:1088–1094. https://doi.org/10.1016/j.jbiomech.2009.02.030

[12] Ruimerman R, Hilbers P, Van Rietbergen B. & Huiskes R. A theoretical framework for strain-related trabecular bone maintenance and adaptation. Journal of biomechanics 2005;38;4:931–941. https://doi.org/10.1016/j.jbiomech.2004.03.037[13] Fernandes P, Rodrigues H & Jacobs C. A model of bone adaptation using a global optimisation criterion based on the trajectorial theory of Wolff. Computer Methods in Biomechanics and Biomedical Engineering 1999;2:125–138. https://doi.org/10.1080/10255849908907982

[14] Van Oers RFM, Ruimerman R, Tanck E, Hilbers PAJ, Huiskes R. A unified theory for osteonal and hemi-osteonal remodeling. Bone 2008;42;2:9–250. https://doi.org/10.1016/j.bone.2007.10.009

[15] Vajda EG, Bloebaum RD. Age-related hypermineralization in the female proximal human femur. Anat Rec. 1999;255:202–11. https://doi.org/10.1002/(SICI)1097-0185(19990601)255:2%3C202::AID-AR10%3E3.0.CO;2-0

[16] Meta M. Young-elderly differences in bone density, geometry and strength indices depend on proximal femur sub-region: A cross sectional study in Caucasian-American women. Bone. 2006;39;1: 152–158. https://doi.org/10.1016/j.bone.2005.11.020

[17] Hadjidakis DJ, Androulakis II. Bone remodeling. Ann N Y Acad Sci. 2006;1092:385-96. https://doi.org/10.1196/annals.1365.035[18] Burr DB, Martin RB, Schaffler MB, Radin EL. Bone remodeling in response to in vivo fatigue microdamage. J Biomech. 1985;18:189–200. https://doi.org/10.1016/0021-9290(85)90204-0

[19] Burr DB, Milgrom C, Boyd RD, Higgins WL, Robin G, Radin EL. Experimental stress fractures of the tibia. Biological and mechanical aetiology in rabbits. Journal of Bone and Joint Surgery. 1990;72B:370–375. https://doi.org/10.1302/0301-620X.72B3.2341429

[20] Burr DB, Martin RB. Calculating the probability that microcracks initiate resorption spaces. J Biomech. 1993;26:613–616. https://doi.org/10.1016/0021-9290(93)90023-8

[21] Burr DB. Bone material properties and mineral matrix contributions to fracture risk or age in women and men. J Musculoskel Neuron Interact 2002; 2;3:201-204.

[22] Parfitt AM. Targeted and nontargeted bone remodeling: relationship to basic multicellular unit origination and progression. Bone 2002;30;1:5–7. https://doi.org/10.1016/s8756-3282(01)00642-1

[23] Mullender MG, Huiskes R. Osteocytes and bone lining cells: Which are the best candidates for mechano-sensors in cancellous bone? Bone 1997;20;6:527-532. https://doi.org/10.1016/s8756-3282(97)00036-7

[24] García-Aznar JM, Rueberg T, Doblare M. A bone remodelling model coupling microdamage growth and repair by 3D BMU-activity. Biomechanics and Modeling in Mechanobiology, 2005;4;2-3: 147-167. https://doi.org/10.1007/s10237-005-0067-x

[25] Donahue SW, Galley SA. Microdamage in bone: implications for fracture, repair, remodeling, and adaptation. Crit Rev Biomed Eng. 2006;34;3:215-71. http://dx.doi.org/10.1615/CritRevBiomedEng.v34.i3.20

[26] Verborgt O, Gibson GJ, Schaffler MB. Loss of osteocyte integrity in association with microdamage and bone remodeling after fatigue in vivo. J Bone Miner Res. 2000;15;1:60-7. https://doi.org/10.1359/jbmr.2000.15.1.60

[27] Bentolila V, Boyce TM, Fyhrie DP, Drumb R, Skerry TM, Schaffler MB. Intracortical remodeling in adult rat long bones after fatigue loading. Bone. 1998;23;3:275-81. https://doi.org/10.1016/s8756-3282(98)00104-5

[28] Wasserman N, Yerramshetty J, Akkus O. Microcracks colocalize within highly mineralized regions of cortical bone tissue. Eur J Morphol. 2005;42;1-2:43-51. https://doi.org/10.1080/09243860500095471

[29] Alan Boyde. The real response of bone to exercise. J Anat. 2003;203;2:173–189. https://doi.org/10.1046/j.1469-7580.2003.00213.x

[30] Eriksen EF, Gundersen HJG, Melsen F & Mosekilde L. Reconstruction of the formative site in iliac trabecular bone in 20 normal individuals employing a kinetic model for matrix and mineral apposition. Metabolic Bone Disease and Related Research 1984;5;5:243–252. https://doi.org/10.1016/0221-8747(84)90066-3

[31] Parfitt AM. The coupling of bone formation to bone resorption : A critical analysis of the concept and of its relevance to the pathogenesis of osteoporosis. Metabolic Bone Disease and Related Research 1982;4;1:1–6.

[32] Currey JD. Effects of differences in mineralization on the mechanical properties of bone. Philos Trans R Soc Lond B Biol Sci. 1984;304;1121:509-18. https://doi.org/10.1098/rstb.1984.0042

[33] Currey JD, Brear K, Zioupos P. The effects of ageing and changes in mineral content in degrading the toughness of human femora. J Biomech. 1996;29;2:257-60. https://doi.org/10.1016/0021-9290(95)00048-8

[34] Diab T, Vashishth D. Morphology, localization and accumulation of in vivo microdamage in human cortical bone. Bone. 2007;40;3:612-8. https://doi.org/10.1016/j.bone.2006.09.027

[35] Mashiba T, Hui S, Turner CH, Mori S, Johnston CC, Burr DB. Bone remodeling at the iliac crest can predict the changes in remodeling dynamics, microdamage accumulation, and mechanical properties in the lumbar vertebrae of dogs. Calcif Tissue Int 2005;77:180–185. https://doi.org/10.1007/s00223-005-1295-x

[36] Herman BC, Cardoso L, Majeska RJ, Jepsen KJ and Schaffler MB. Activation of Bone Remodeling after Fatigue: Differential Response to Linear Microcracks and Diffuse Damage. Bone. 2010 Oct; 47;4:766–772. https://doi.org/10.1016/j.bone.2010.07.006

[37] Garnero P. The contribution of collagen crosslinks to bone strength. Bonekey Rep. 2012; 1: 182. https://doi.org/10.1038/bonekey.2012.182

[38] Henriksen K, Leeming DJ, Byrjalsen I, Nielsen RH, Sorensen MG, Dziegiel MH, John Martin T, Christiansen C, Qvist P, Karsdal MA. Osteoclasts prefer aged bone. Osteoporos Int (2007) 18:751–759. https://doi.org/10.1007/s00198-006-0298-4

[39] Krukowski M, Kahn AJ. Inductive specificity of mineralized bone matrix in ectopic osteoclast differentiation. Calcif Tissue Int. 1982 Sep;34(5):474-9. https://doi.org/10.1007/bf02411288

[40] Boyde A, Jones SJ (1987). Early scanning electron microscopic studies of hard tissue resorption: their relation to current concepts reviewed. Scanning Microsc vol. 1, (1) 369-381.

[41] Glowacki J, Rey C, Cox K, Lian J. Effects of bone matrix components on osteoclast differentiation. Connect Tissue Res. 1989;20(1-4):121-9. https://doi.org/10.3109/03008208909023880[42] Turner CH, Rho J, Takano Y, Tsui TY, Pharr GM. The elastic properties of trabecular and cortical bone tissues are similar: results from two microscopic measurement techniques. Journal of Biomechanics 1999;32:437-441. https://doi.org/10.1016/s0021-9290(98)00177-8

[43] Odgaard A. Three-dimensional methods for quantification of cancellous bone architecture. Bone 1997;20: 315-28. https://doi.org/10.1016/S8756-3282(97)00007-0.

[44] Harrigan TP, Mann RW. Characterization of microstructural anisotropy in orthotropic materials using a second rank tensor. J Mater Sci. 1984;19: 761-767. https://doi.org/10.1007/BF00540446

[45] Homminga J, Huiskes R, Van Rietbergen B, Rüegsegger P, Weinans H. Introduction and evaluation of a gray-value voxel conversion technique. Journal of Biomechanics 2001;34:513-517. https://doi.org/10.1016/s0021-9290(00)00227-x

[46] Ulrich D, Van Rietbergen B, Weinans H, Rüegsegger P. Finite element analysis of trabecular bone structure: a comparison of image-based meshing techniques. Journal of Biomechanics 1998;31:1187-1192. https://doi.org/10.1016/s0021-9290(98)00118-3

[47] Hoc T, Henry L, Verdier M, Aubry D, Sedel L & Meunier A. Effect of microstructure on the mechanical properties of haversian cortical bone. Bone 2006;38;4:466–474. https://doi.org/10.1016/j.bone.2005.09.017

[48] Zebaze RMD, Jones AC, Pandy MG, Knackstedt MA & Seeman E. Differences in the degree of bone tissue mineralization account for little of the differences in tissue elastic properties. Bone 2011;48:1246–1251. https://doi.org/10.1016/j.bone.2011.02.023

[49] Garnero P, Sornay-Rendu E, Chapuy MC & Delmas PD. Increased bone turnover in late postmenopausal women is a major determinant of osteoporosis. Journal of Bone and Mineral Research 1996;11;3:337–349. https://doi.org/10.1002/jbmr.5650110307

[50] Morgan EF, Bayraktar HH, Keaveny TM. Trabecular bone modulus-density relationships depend on anatomic site. J Biomech. 2003 Jul; 36;7:897-904. https://doi.org/10.1016/s0021-9290(03)00071-x

[51] Recker RR, Kimmel DB, Parfitt AM, Davies KM, Keshawarz N, Hinders S. Static and tetracycline-based bone histomorphometric data from 34 normal postmenopausal females. J Bone Miner Res. 1988;3;2:133-44. https://doi.org/10.1002/jbmr.5650030203

[52] Riggs BL, Melton LJ, Robb RA, Camp JJ, Atkinson EJ, Oberg AL, Rouleau PA, McCollough CH, Khosla S and Bouxsein ML. Population-Based Analysis of the Relationship of Whole BoneStrength Indices and Fall-Related Loads to Age- and Sex-Specific Patterns of Hip and Wrist Fractures. J Bone Miner Res. 2006;21;2:315-23. https://doi.org/10.1359/JBMR.051022

[53] Yuen KW, Kwok TC, Qin L, Leung JC, Chan DC, Kwok AW, Woo J, Leung PC. Characteristics of age-related changes in bone compared between male and female reference Chinese populations in Hong Kong: a pQCT study. J Bone Miner Metab. 2010;28;6:672-81. https://doi.org/10.1007/s00774-010-0170-7

[54] Macdonald HM, Nishiyama KK, Kang J, Hanley DA, Boyd SK. Age-related patterns of trabecular and cortical bone loss differ between sexes and skeletal sites: A population-based HR-pQCT study. Journal of Bone and Mineral Research, 2011;26;1:50–62. https://doi.org/10.1002/jbmr.171

[55] Chen H, Zhou X, Fujita H, Onozuka M, Kubo KY. Age-Related Changes in Trabecular and Cortical Bone Microstructure. Int J Endocrinol. 2013;2013:213234. http://dx.doi.org/10.1155/2013/213234

[56] Mazess RB. On aging bone loss. Clinical Orthopaedics and Related Research 1982;165:239–252.

[57] Kostenuik PJ, Capparelli C, Morony S, Adamu S, Shimamoto G, Shen V, Lacey DL, Dunstan CR. OPG and PTH-(1–34) Have Additive Effects on Bone Density and Mechanical Strength in Osteopenic Ovariectomized Rats. Endocrinology, October 2001, 142;10:4295-4304. https://doi.org/10.1210/endo.142.10.8437

[58] Shahnazari M, Yao W, Wang B, Panganiban B, Ritchie RO, Hagar Y, Lane NE. Differential Maintenance of Cortical and Cancellous Bone Strength Following Discontinuation of Bone-Active Agents. J Bone Miner Res. 2011 Mar; 26;3: 569–581. https://doi.org/10.1002/jbmr.249

[59] Chavassieux P, Meunier PJ, Roux JP, Portero-Muzy N, Pierre M, Chapurlat R. Bone Histomorphometry of Transiliac Paired Bone Biopsies After 6 or 12 Months of Treatment With Oral Strontium Ranelate in 387 Osteoporotic Women: Randomized Comparison to Alendronate. JBMR, 2014. Vol29(3):618-28. https://doi.org/10.1002/jbmr.2074

[60] Dufresne TE, Chmielewski PA, Manhart MD, Johnson TD, Borah B. Risedronate preserves bone architecture in early postmenopausal women in 1 year as measured by three-dimensional microcomputed tomography. Calcif Tissue Int. 2003 Nov; 73(5):423-32. https://doi.org/10.1007/s00223-002-2104-4

[61] Khajuria DK, Razdan R, Mahapatra DR. Zoledronic acid in combination with alfacalcidol has additive effects on trabecular microarchitecture and mechanical properties in osteopenic ovariectomized rats. J Orthop Sci. 2014;19;4:646-56. https://doi.org/10.1007/s00776-014-0557-8

[62] Keller TS. Predicting the compressive mechanical behavior of bone. J Biomech. 1994;27;9:1159-68. https://doi.org/10.1016/0021-9290(94)90056-6

[63] Bone HG, Hosking D, Devogelaer JP, Tucci JR, Emkey RD, Tonino RP, Rodriguez-Portales JA, Downs RW, Gupta J, Santora AC, Liberman UA. Ten years' experience with alendronate for osteoporosis in postmenopausal women. N Engl J Med 2004;350:1189–1199. https://doi.org/10.1056/NEJMoa030897

[64] Cheng Z, Yao W, Zimmermann EA, Busse C, Ritchie RO, Lane NE. Prolonged Treatments with Antiresorptive Agents and PTH Have Different Effects on Bone Strength and the Degree of Mineralization in Old Estrogen-Deficient Osteoporotic Rats. J Bone Miner Res. 2009;24;2:209–220. https://doi.org/10.1359/jbmr.81005

[65] Li J, Mashiba T, Burr DB. Bisphosphonate treatment suppresses targeted repair of microdamage. Trans Orthop Res Soc. 2001;26:320. https://doi.org/10.1007/s002230010036

[66] Graeff C, Chevalier Y, Charlebois M, Varga P, Pahr D, Nickelsen TN, Morlock MM, Gluer CC and Zysset PK. Improvements in Vertebral Body Strength Under Teriparatide Treatment Assessed In Vivo by Finite Element Analysis: Results From the EUROFORS. J Bone Miner Res. 2009;24;10:1672-80. https://doi.org/10.1359/jbmr.090416

[67] Dempster DW, Cosman F, Kurland ES, Zhou H, Nieves J, Woelfert L, Shane E, Plavetić K, Müller R, Bilezikian J, Lindsay R. Effects of Daily Treatment with Parathyroid Hormone on Bone Microarchitecture and Turnover in Patients with Osteoporosis: A Paired Biopsy Study. 2001, Vol16(10):1846-1853. https://doi.org/10.1359/jbmr.2001.16.10.1846

[68] Ribot C, Trémollières F. Place du traitement hormonal substitutif dans la prise en charge de l'ostéoporose postménopausique et la prévention du risque fracturaire. Presse Med. 2006;35;10;2:1557-6. https://doi.org/10.1016/S0755-4982(06)74851-5

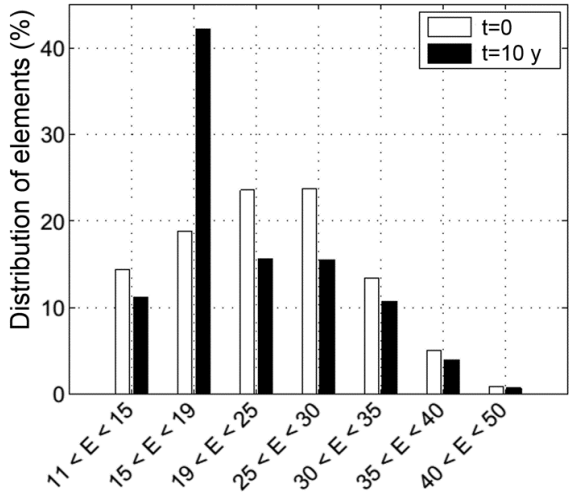 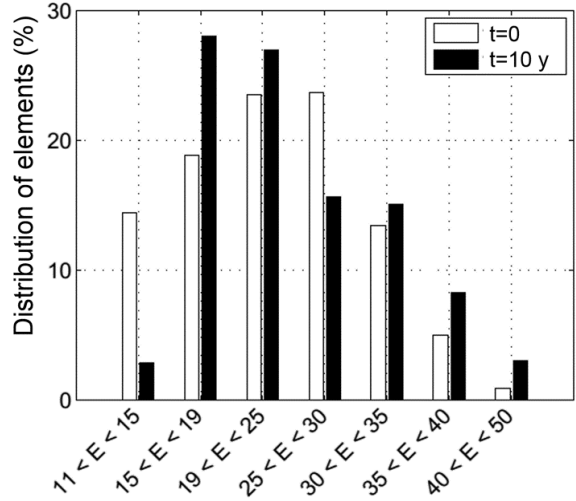

$K_{hyper} = 0,1\text{GPa}, \alpha = 46,7\%$  $K_{hyper} = 0,3\text{GPa}, \alpha = 69,0\%$

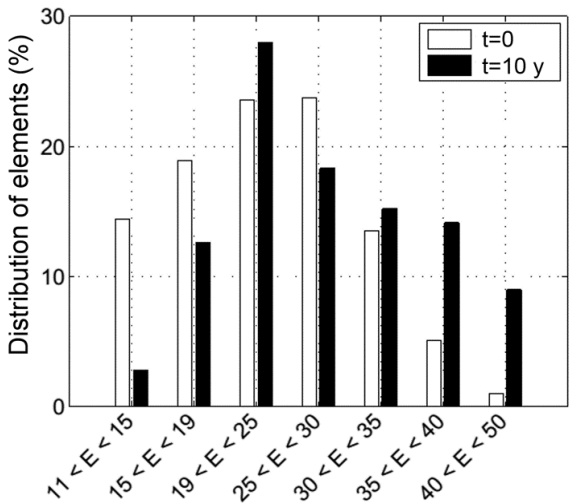 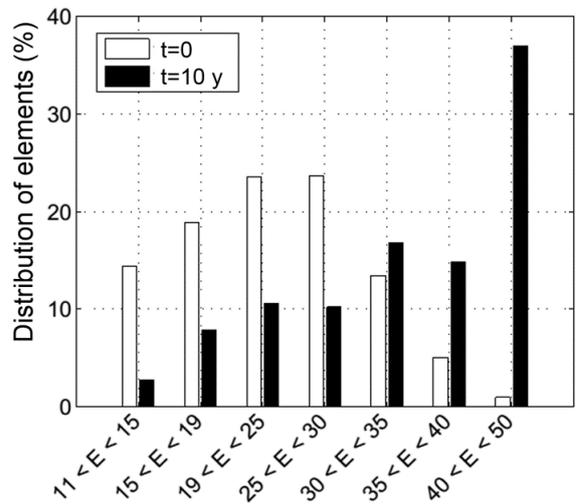

$K_{hyper} = 0,5\text{GPa}, \alpha = 84,6\%$  $K_{hyper} = 1\text{GPa}, \alpha = 89,4\%$

Figure 1 - Number of elements according to Young's modulus, before remodeling (white) and after 10 years of remodeling (black) for different $K_{hyper}$ values. α is the proportion of hypermineralized elements (E ≥ 19GPa), given that for t = 0, α = 66.7%.

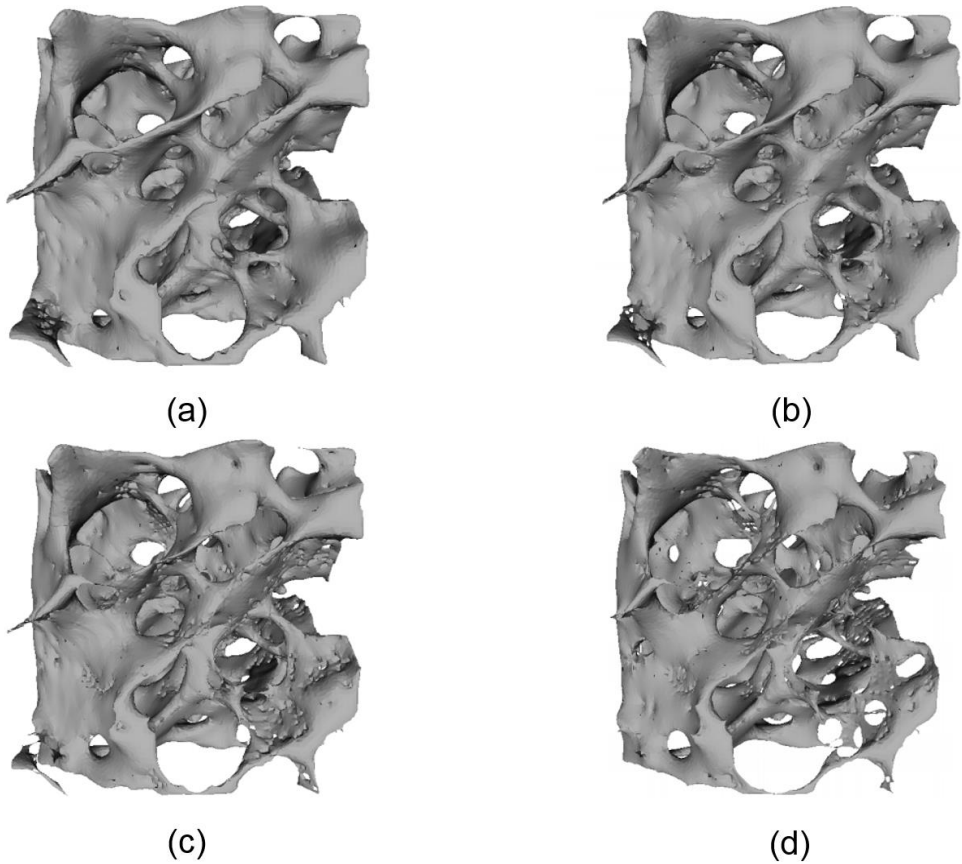

Figure 2 - Evolution of the microstructure shown at the initial state (a) and after 20 years of remodeling for different pairs of parameters RA and δ: (b) RA = 7%, δ =-5%, (c) RA = 14%, δ =-5% and (d) RA = 14%, δ =-10%. The microstructures were smoothed after remodeling with MeshLab (MeshLab Visual Computing Lab - ISTI – CNR)

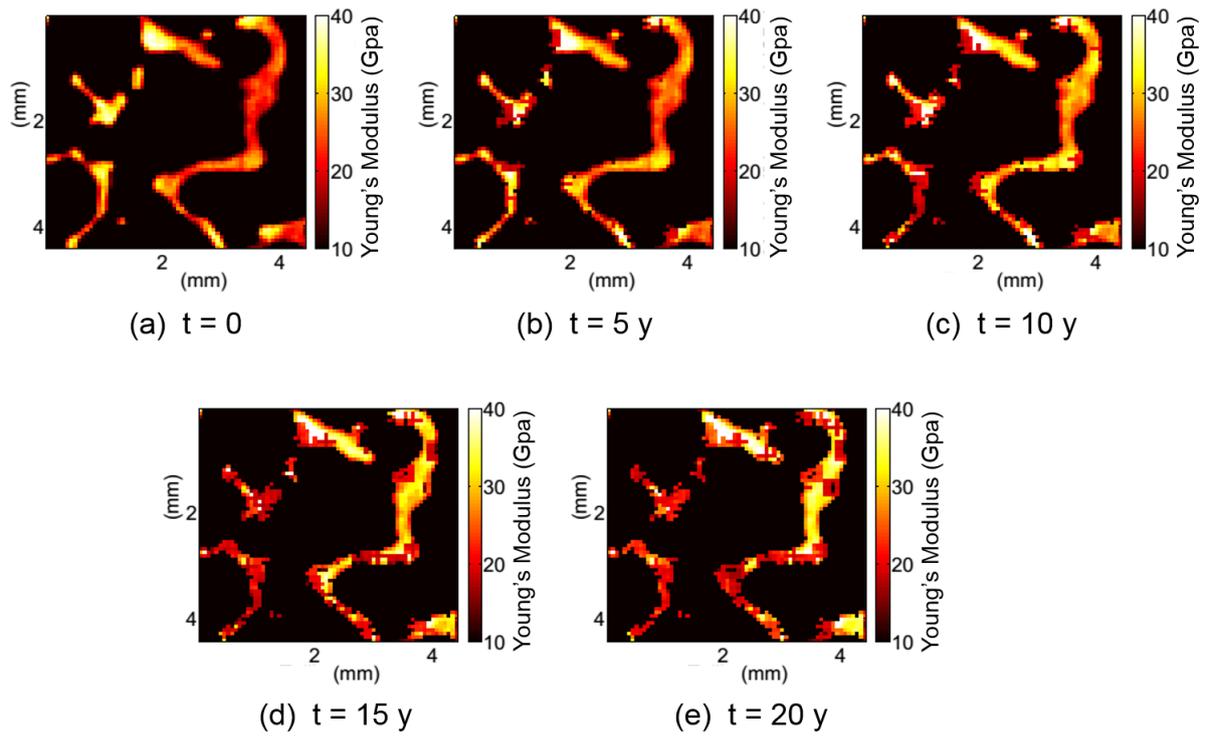

Figure 3 - Evolution of local Young's modulus in cross-section view after 5, 10, 15 and 20 years of remodeling in the normal aging case

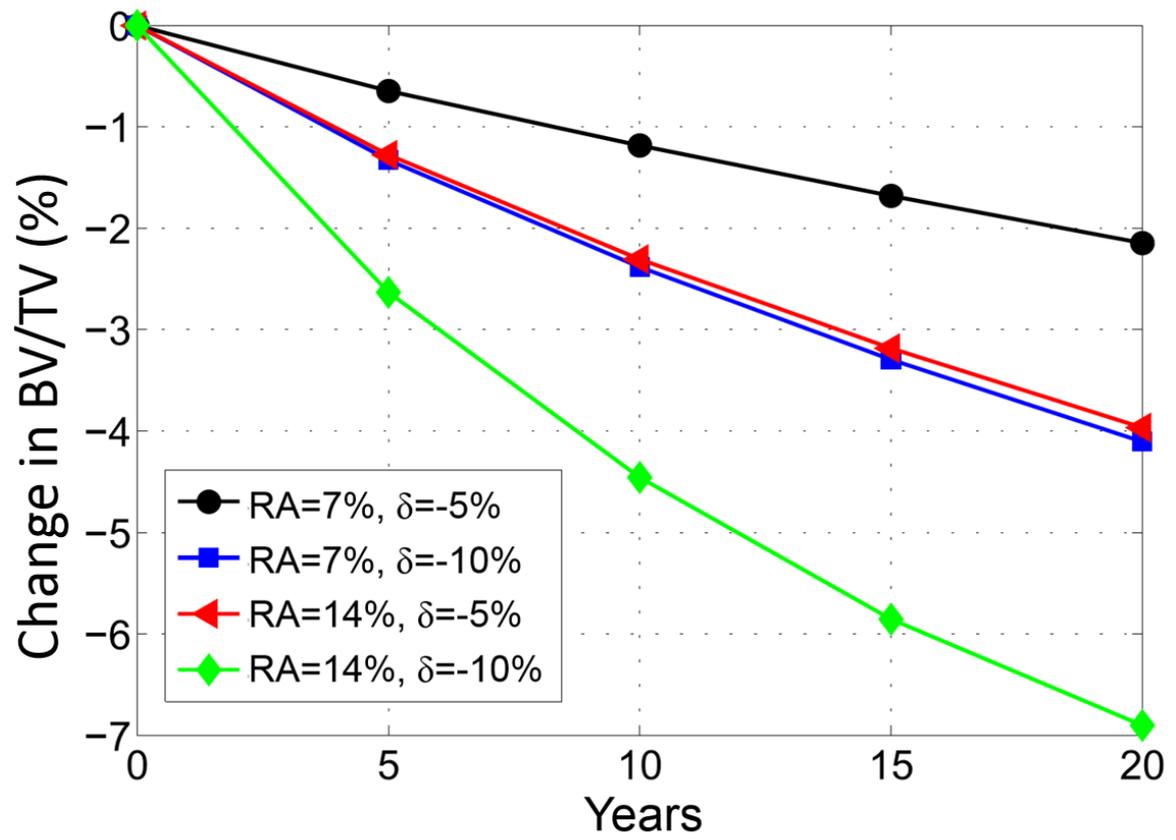

Figure 4 - Evolution of the ratio BV/TV for different pairs of parameters (RA, δ) representing healthy or pathologic cases

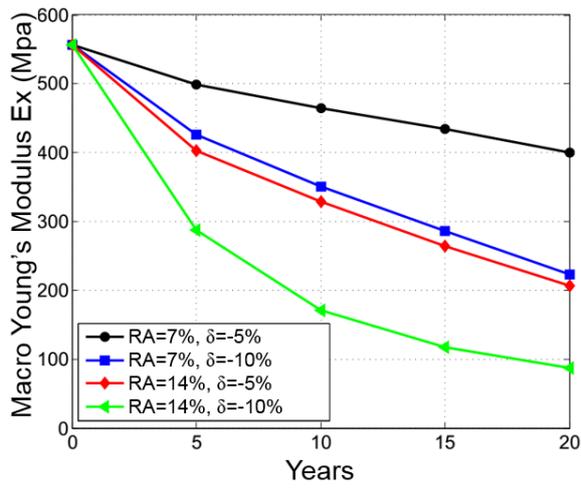
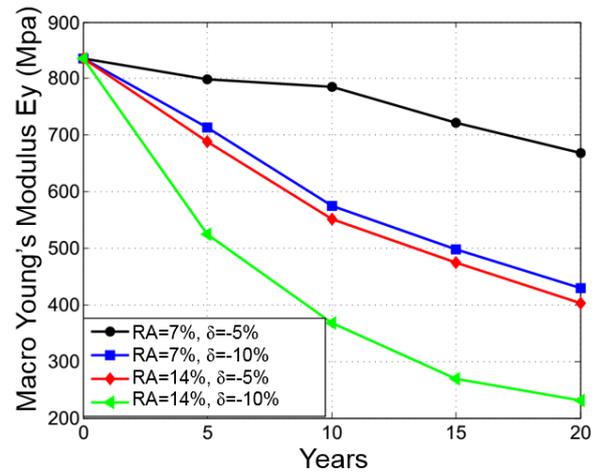
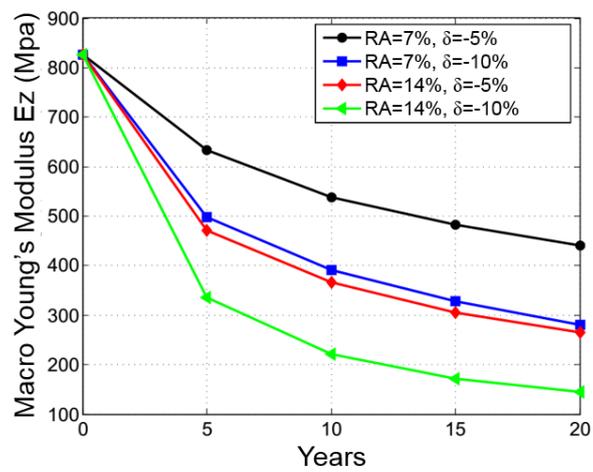

Figure 5 - Evolutions of the apparent Young Modulus (Ex, Ey and Ez) after 20 years of remodeling for different pairs of parameters RA and δ.

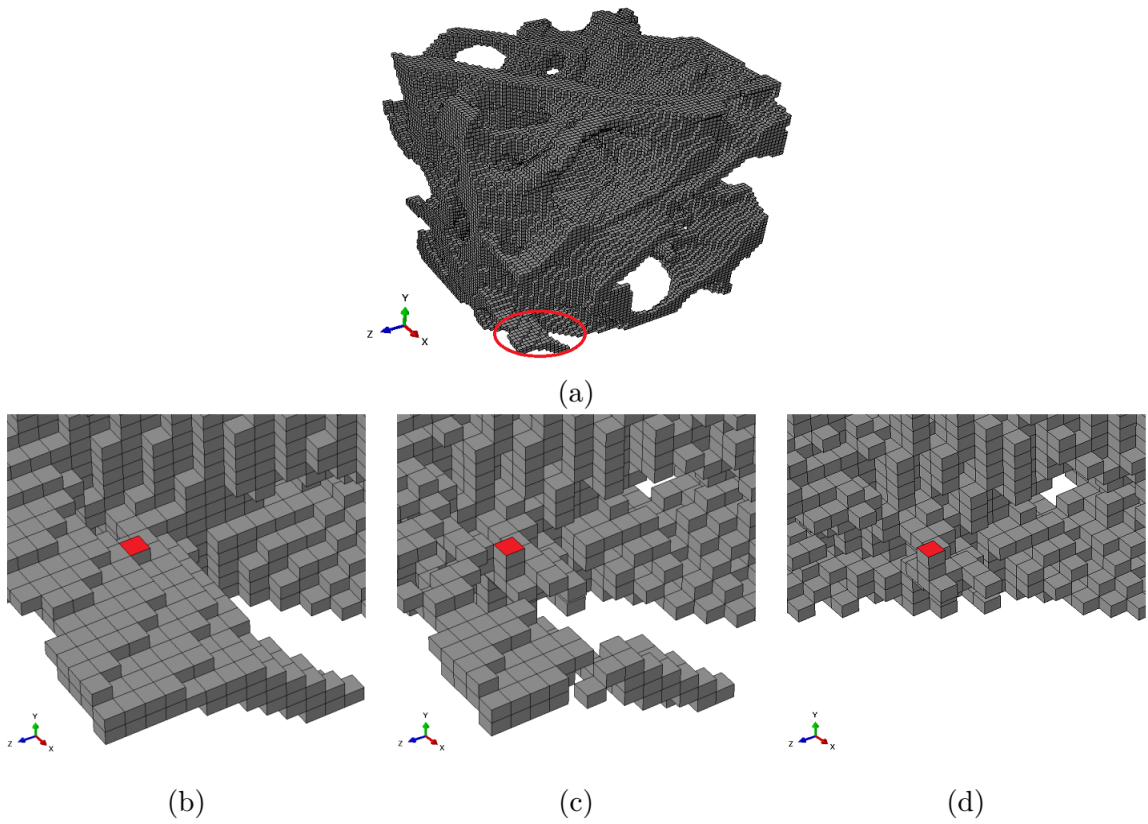

Figure 6 – Mode of degradation of bone structure. Initial microstructure (a), close-up of circled part (b), perforation of the trabecular plate after 10 years of remodeling (c) and complete removal of part after 15 years of remodeling (d). Simulation parameters: RA = 14% and δ = -10%.

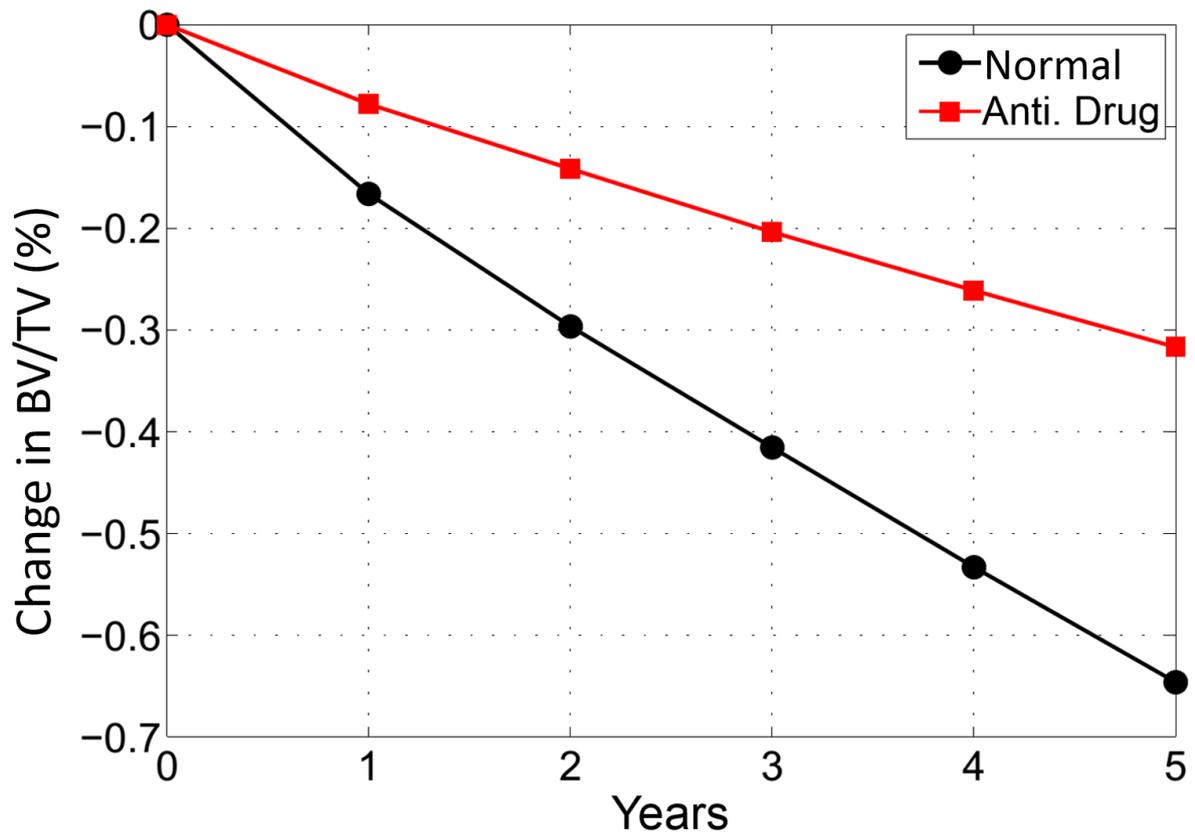

Figure 7 - Reduction of bone loss in antiresorptive therapy (RA = 3.5%, δ = -5%) over 5 years compared to healthy aging (RA = 7%, δ = -5%). Note that total volume (TV) does not change over time, and that relative change in BV/TV= relative change in BV.

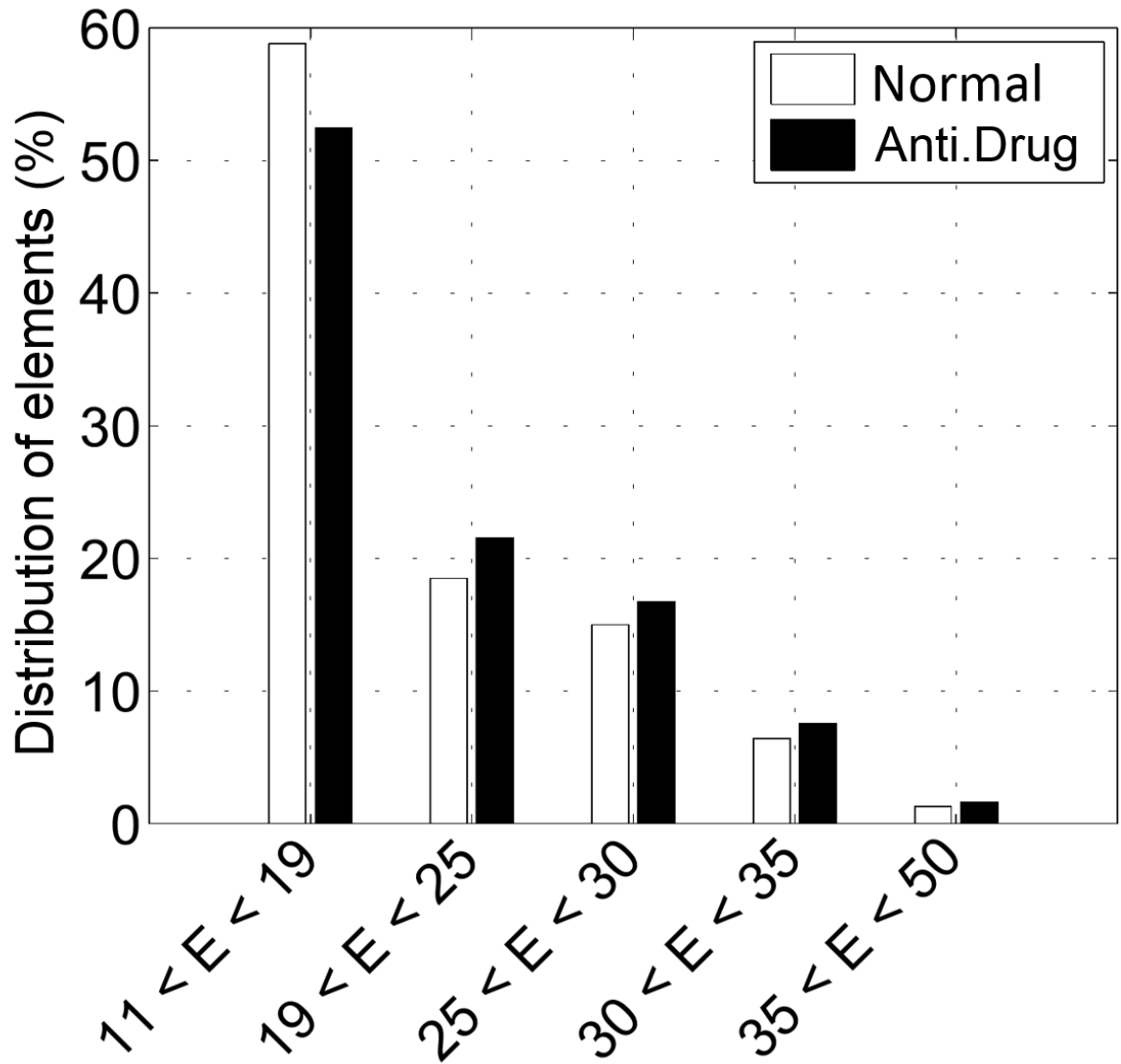

Figure 8 - Proportion of elements according to their local Young's modulus (GPa) after 5 years of antiresorptive treatment compared to normal aging. After 5-year antiresorptive treatment, 47.5% of bone elements are hypermineralized (E ≥ Ecort); they are 41.2% after 5-year normal aging.

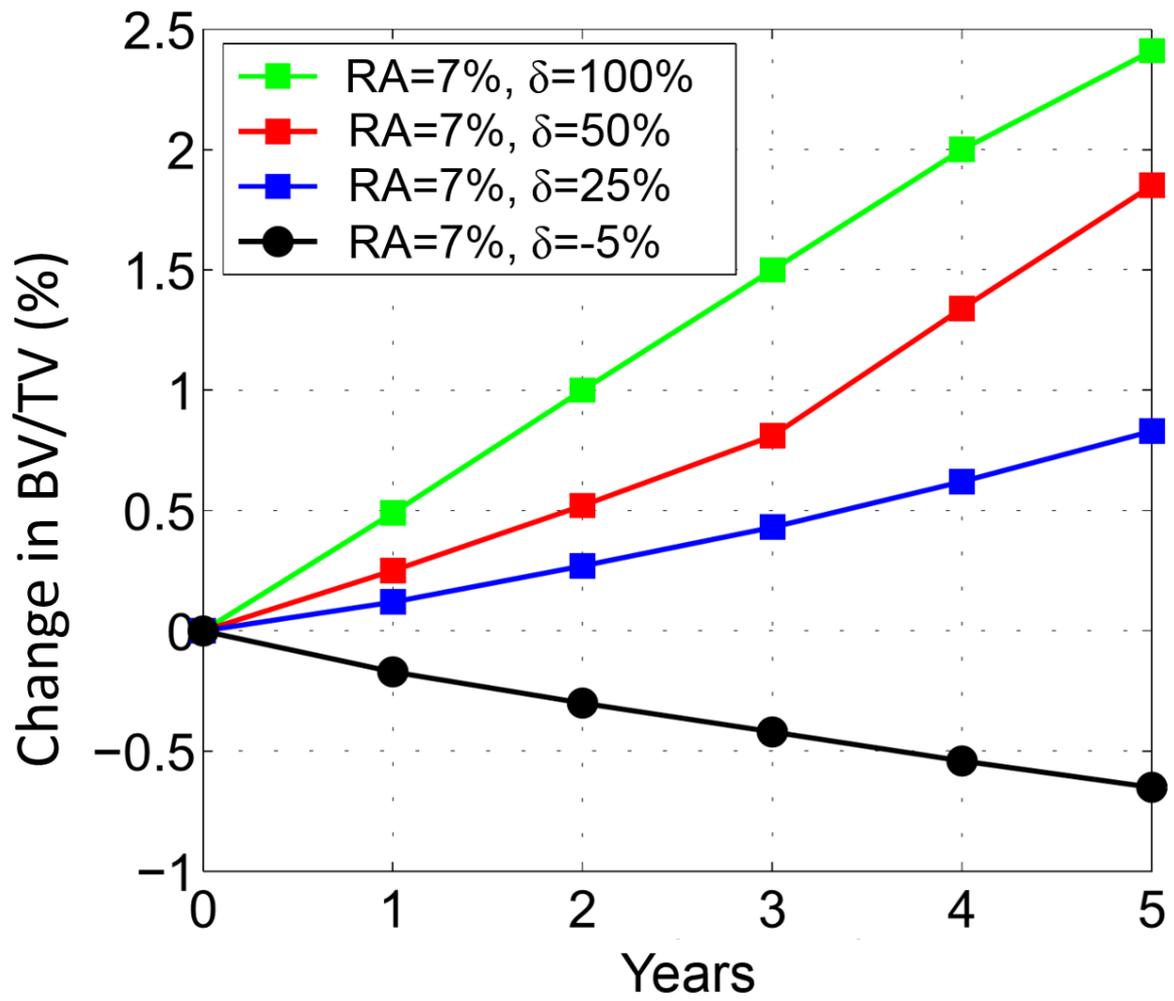

Figure 9 - Comparison of the evolution of the bone fraction under anabolic treatments (δ ≥ 0) compared to a healthy case.

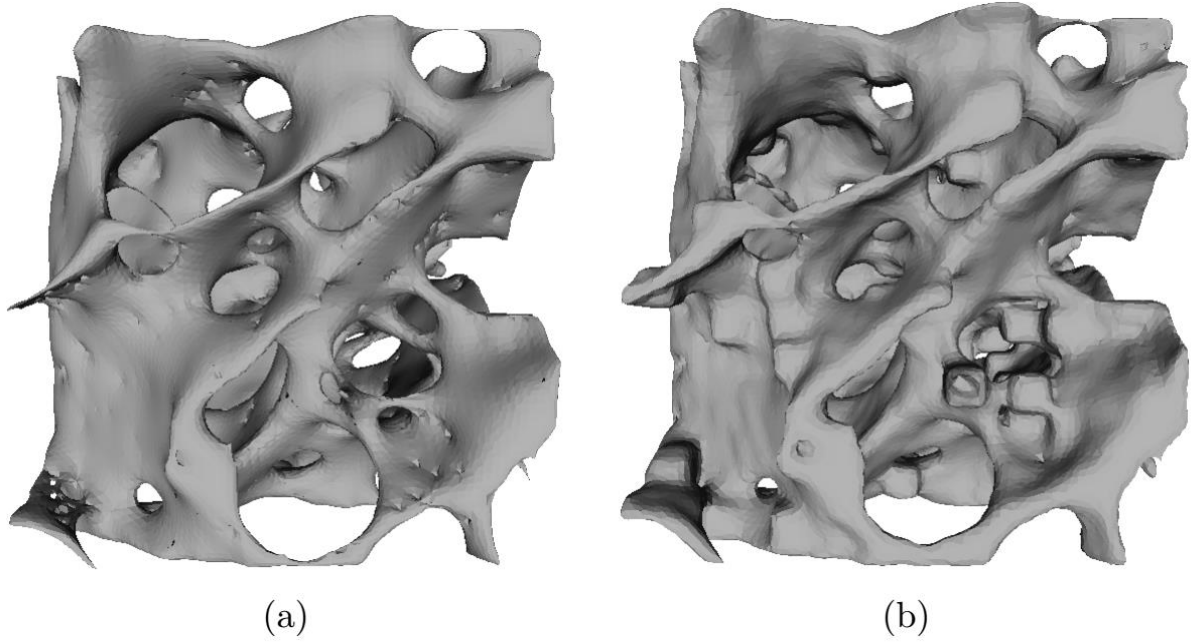

Figure 10 – Results of simulations of anabolic treatments on the thickening of trabecular spans and plates. (a) before and (b) after 5 years of treatment with an anabolic (δ = 50%).

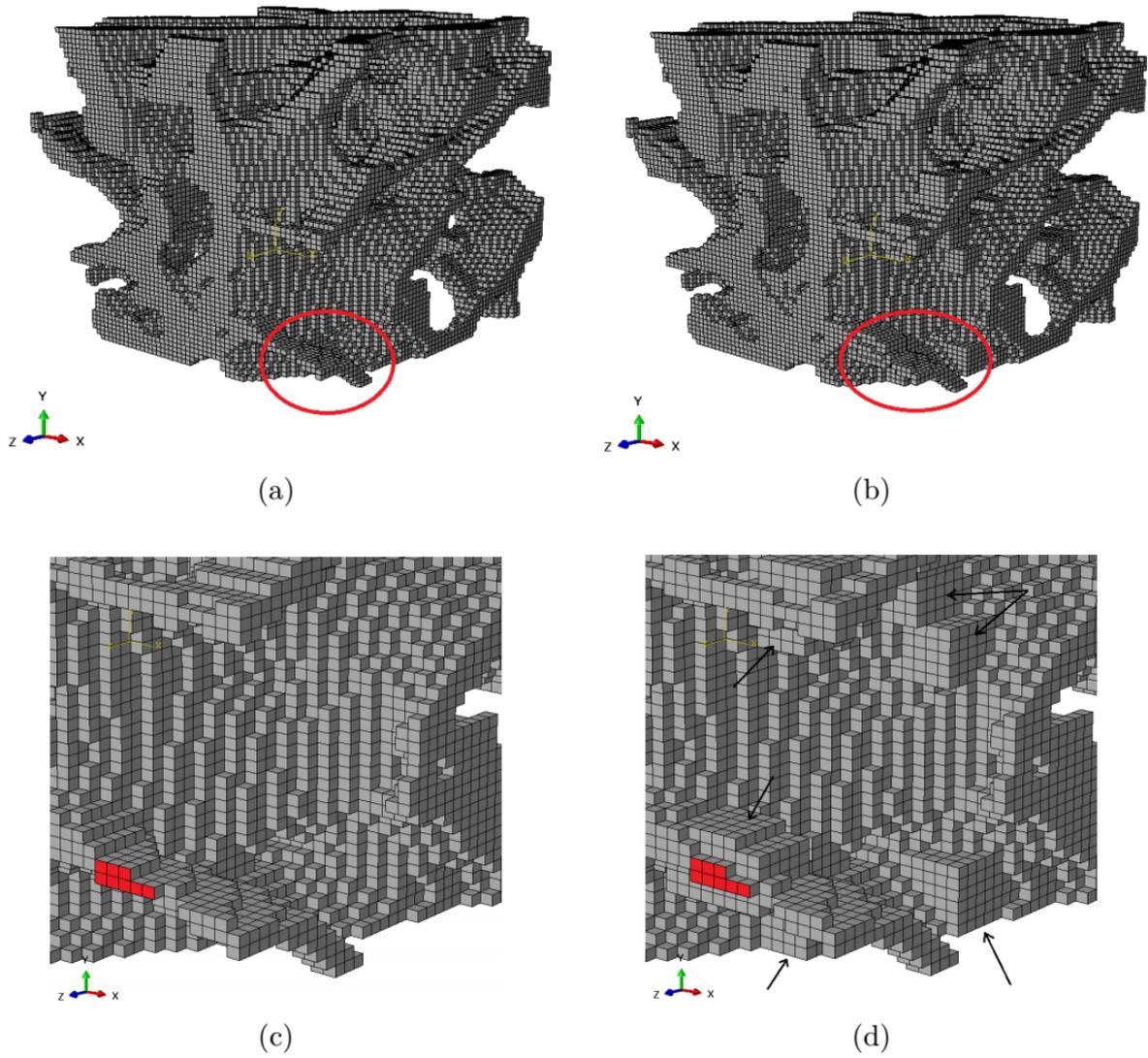

Figure 11 - Voxel geometries before (a) and (b) after 5 years of treatment with an anabolic (δ =50%). Close-ups (c) and (d) of the circles highlighting the areas of bone gain (arrows).

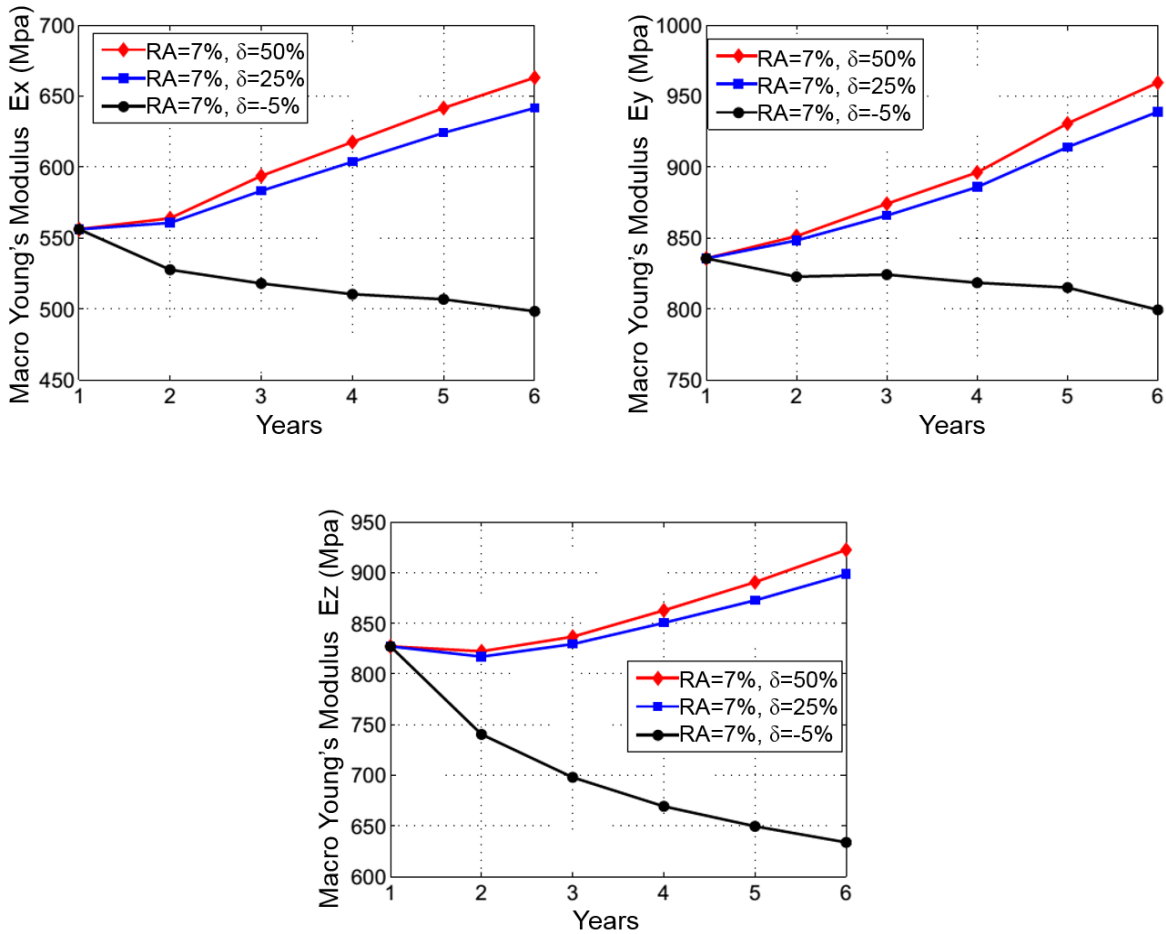

Figure 12 - Evolutions of equivalent macro Young's modulus after 5 years of anabolic treatment (RA = 7% ; δ =25%) and (RA = 7% ; δ =50%) compared to healthy aging (RA = 7% ; δ =-5%). Note that axis scales are different from scales of Figure 5.

| Bone remodeling cases | Remodeling Area (RA) | Resorpt. Vs deposit. Imbalance (δ) |
|---|---|---|
| Normal aging | 7% | -5% |
| Osteoporosis | 7 ; 14% | -10 ; -5% |
| Antiresorptive treatment | 3.5% | -5% |
| Anabolic treatment | 7 % | 25 ; 50 ; 100% |

Table 1 - Range of parameter values used for the different simulations

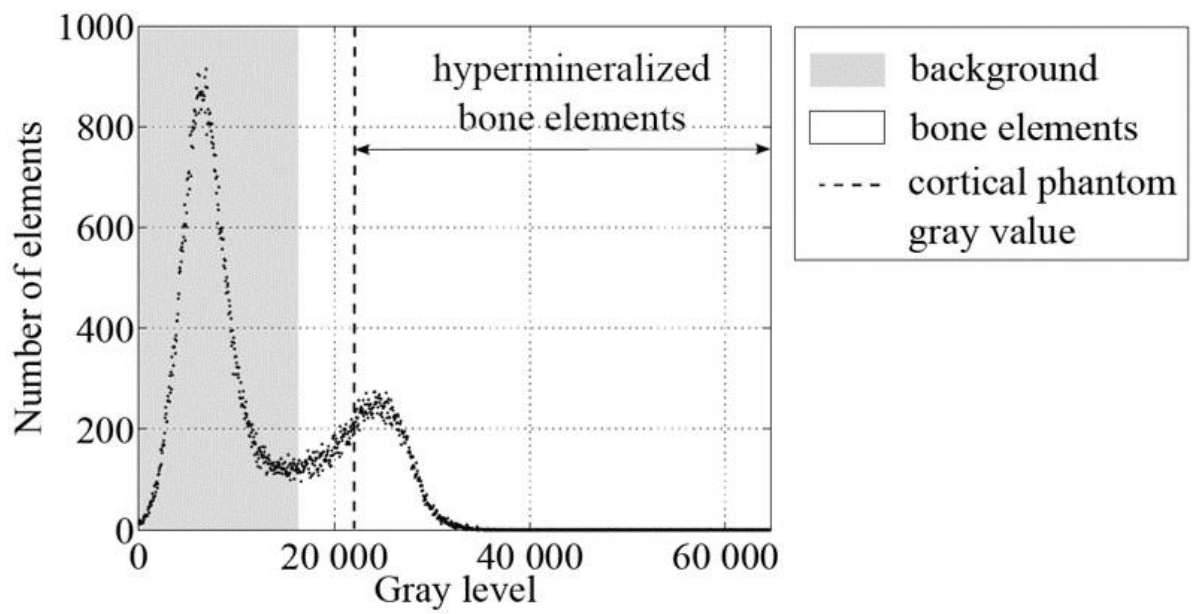

Supplementary Data Figure 1 - Grayscale distribution after µCT acquisition: vertical black-dash line corresponds to the cortical phantom gray-value

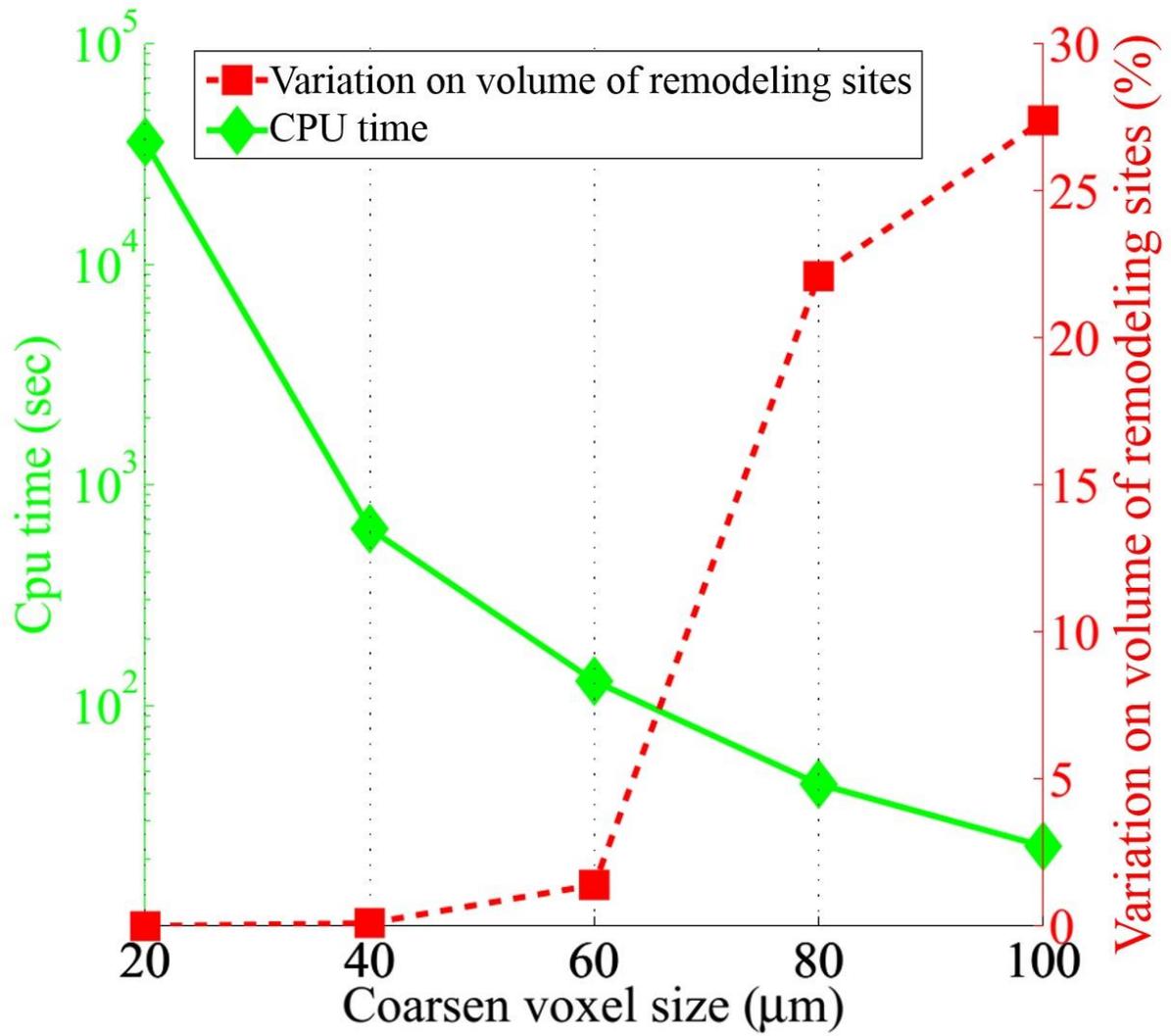

Supplementary Data Figure 2 - Number of elements and mean variation on Young modulus (a) - cpu time and variation on volume of remodeling sites (b) for coarsened models

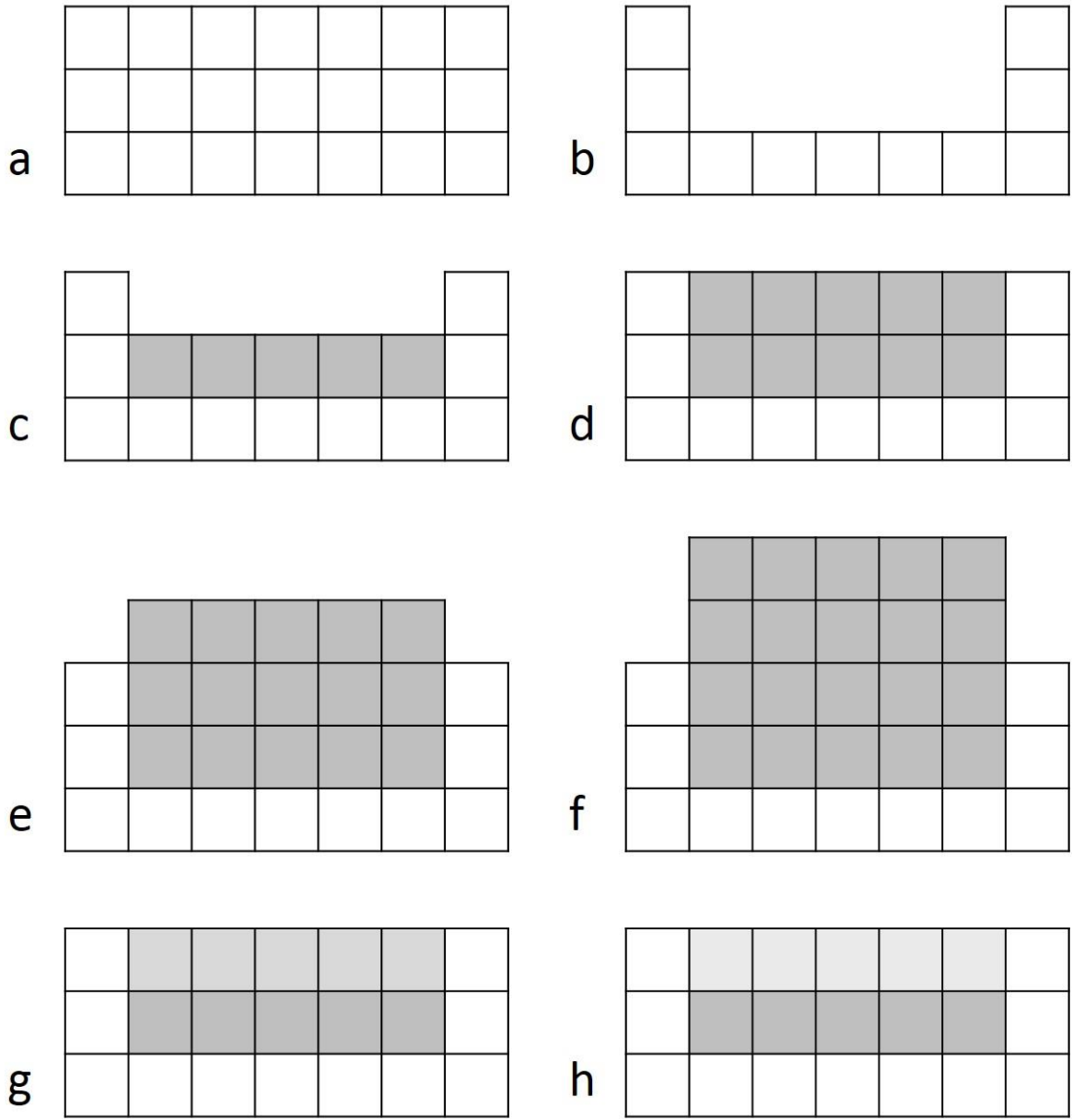

Supplementary Data Figure 3 - Simulation of bone volume change at a remodeling site. a) initial state of bone matrix; b) after resorption, two layers of elements have been removed; c) after apposition with δ = -50%, there is a net loss of one layer of elements; d) after apposition with δ = 0%, there is no net gain in bone volume, the number of elements has not changed but their density is lower, e) after apposition with δ = 50%, there is a net gain in bone volume of one element layer; f) after apposition with δ = 100%, there is a net gain in bone volume of two element layers. g) after apposition with δ =-25%, elements of the lower layer at a remodeling site have been given the density and the Young's modulus of cortical bone ($E_{cort}$=19 GPa) while elements of the top layer have been given a density of 0.5 and a Young's modulus of 9.5 GPa. h) If the site in g) was remodeled again immediately, without additional mineralization, with the same bone balance of δ =-25%, the elements of top layer would be given a density = 0.125 and a Young's modulus of 2.4 GPa while elements of the lower layer at a remodeling site would be given the density and the Young's modulus of cortical bone ($E_{cort}$).

- *BS/BV=10.18mm$^{-1}$,*
- *14% of surface is remodeled every year (y)*
- *Resorption pit is a 0.150mm-radius half sphere*
- *Osteoblasts deposit 95% of the volume resorbed*

The rate of Bone Formation/Bone Volume/year (BF/BV/y)

$= 0.95*$Bone Resorption/BV/y

$= 0.95*$site number/y $*$ pit volume /BV

$= 0.95* 0.14*$BS/pit area $*$ pit volume / BV

$= 0.95* 0.14*$BS$/\pi/0.15^2 * 1/2*4/3*\pi*0.15^3$ / BV

$= 0.95* 0.14*4/3*0.15*1/2*$BS/BV

$= 0.95*0.14*4/3*0.15*1/2*10.18$

$= 13{,}5\ \%$

Supplementary Data Figure 4 – Calculation of the rate of Bone Formation/Bone Volume/year (BF/BV/y) under simulated normal aging of trabecular bone